\documentclass[twocolumn,aps,pra]{revtex4}
\usepackage{epsfig}
\usepackage[english]{babel}
\usepackage{latexsym}
\usepackage{graphics}
\usepackage{subfigure}
\usepackage{graphicx}
\usepackage{dcolumn}
\usepackage{amsmath}
\usepackage{hyperref}
\usepackage{amssymb}
\usepackage{color}
\usepackage{caption3}

\begin{document}

\title{Response time of photoemission at quantum-classic boundary}

\author{J. Y. Che$^{1,\dag}$, C. Chen$^{1,\dag}$, W. Y. Li$^{2}$, S. Wang$^{3}$, X. J. Xie$^{1}$, J. Y. Huang$^{1}$, Y. G. Peng$^{1}$,  G. G. Xin$^{4}$, and Y. J. Chen$^{1,*}$}

\date{\today}

\begin{abstract}

The response time of the electron to light in photoemission is difficult to define and measure.
Tunneling ionization of atoms, a strong-laser-induced photoemission process, provides a semiclassical case for visiting the problem.
Here, we show that the response time can be determined at the boundary between quantum and classic.
Specifically, tunneling is instantaneous but a finite response time (about 100 attoseconds) is needed for the state of the tunneling electron to evolve into the ionized state around tunnel exit.
This time can be well described with a compact expression related to some basic laser and atomic parameters.
Moreover, it can be directly mapped to and easily decoded from photoelectron momentum with a simple mapping, allowing an unambiguous measurement.
These results shed light on definition and measurement of the response time of photoemission.

\end{abstract}

\affiliation{1.College of Physics and Information Technology, Shaan'xi Normal University, Xi'an, China\\
2.School of Mathematics and Science, Hebei GEO University, Shijiazhuang, China\\
3.College of Physics and Hebei Key Laboratory of Photophysics Research and Application, Hebei Normal University, Shijiazhuang, China\\
4.School of Physics, Northwest University, Xi'an, China}
\maketitle

\emph{Introduction}.-At the beginning of last century, Einstein's light quantum hypothesis gives a good explanation for energy-domain law of photoemission,
but the time-domain property of the effect is not discussed in detail.
The accurate measurement and description of the response time of the electron to light in photoemission is very difficult.
Experimentally, direct measurement of this time is still not possible at present, while indirect measurement needs theoretical support \cite{Schultze,Pazourek}.
Theoretically, because there is no time operator in quantum mechanics, it is difficult to define the response time \cite{Maquet1,Saalmann}.
Recently, the development of intense ultrashort laser technology provides the possibility for probing the electronic motion in strong laser-atom interaction with attosecond time resolution
\cite{Krausz2009,Krausz,Maquet,Vrakking}.
In intense laser fields, electron dynamics can be described by semiclassical theory, in which time is easier to define.
One therefore may ask whether the response time of tunneling ionization \cite{Becker2002}, a strong-laser-induced photoemission process, can be probed with the present ultrafast laser technology.

This tunneling-related response time implies the time of strong three-body interaction between laser, electron and nucleus (Coulomb).
Experimentally, because this response time can not be probed directly, a definite mapping  between the observable (e.g., photoelectron momentum) and
this  time is needed as a time-decoding tool.
The treatment of the response time problem therefore converts into finding
this mapping.
The well-known classical or quantum electron-trajectory theory \cite{Becker2002} arising from the simple-man model (SM) \cite{Yang1993,Corkum} or strong-field approximation (SFA) \cite{Lewenstein1994}
provides mappings between time and observable.
But these mappings are based on an assumption neglecting the crucial Coulomb effect.
Recently,  some progresses were made for Coulomb-included electron trajectory  \cite{MishaY,Goreslavski,yantm2010,Torlina,Klaiber,Teeny}.
Due to the difficulty in analytical treatment of the Coulomb potential in three-body interaction, a unified Coulomb-included mapping between time and observable is not to access yet.

Here, we show that through semiclassically determining a transition state at the boundary between quantum and classic,
the difficulty of Coulomb treatment can be overcome and the response time of tunneling ionization can be probed.
Specifically, after the tunneling electron exits the barrier, it is located at a transition state which 
possesses  properties of both bound and continuum states. 
A small period of time is needed for the tunneling electron to evolve from the quasi-bound transition state 
into a Coulomb-free ionized state.
This time reflects the essential response time of the electronic wave function to a tunneling ionization event.
It encodes the main Coulomb effect during tunneling and can be quantitatively described with a compact expression related to
laser intensity, wavelength and atomic ionization energy.
With this expression,
a clear mapping between the response time and the photoelectron momentum can also be established.
The response time determined here is validated by a series of recent tunneling ionization experiments
as the observable deduced from this time 
quantitatively agrees with experiments.

\emph{Theory}.-We begin our discussions with ionization of atoms in strong elliptical laser fields with high ellipticity, as in attoclock experiments   \cite{Eckle1,Eckle2,Eckle3,Undurti,Quan,Landsman,Camus}.
For this elliptical case, some complex effects such as rescattering and quantum interference are negligible
and we can focus on the effect of response time on the photoelectron momentum distribution (PMD).
The elliptical laser field has the electric field  $\mathbf{E}(t)=f(t)[\vec{\mathbf{e}}_{x}E_{x}(t)+\vec{\mathbf{e}}_{y}E_{y}(t)]$,
where $E_{x}(t)=E_0\sin(\omega t)$, $E_{y}(t)=E_1\cos(\omega t)$,
$E_0={E_L}/{\sqrt{1+\epsilon^2}}$ and $E_1=\epsilon {E_L}/{\sqrt{1+\epsilon^2}}$, with $E_L$ being the maximal laser amplitude related to the peak intensity $I$, $\epsilon$  the ellipticity, $\omega$  the laser frequency and $f(t)$  the envelope function.

For strong-field ionization, the mapping relation between the drift momentum $\textbf{p}$ and the ionization time $t_0$ in SM  is $\textbf{p}=-\mathbf{A}(t_0)$.
Here, $\mathbf{A}(t)$ is the vector potential of the electric field $\textbf{E}(t)$.
In SFA, it is $\textbf{p}=\textbf{v}(t_0)-\mathbf{A}(t_0)$ where $\textbf{v}(t_0)$ is the exit velocity of the electron
at the exit position $\textbf{r}(t_0)$ \cite{yantm2010}.
In this paper, with introducing a transition state at the tunnel exit, which contains properties of both bound and continuum states and
satisfies the basic symmetry requirement imposed by the central symmetry of the Coulomb potential,
a mapping between the Coulomb-included momentum $\textbf{p}'$ and  ionization time $t_i$ is constructed (see methods in \cite{method}).
That is
\begin{equation}
\mathbf{p}'=\textbf{v}(t_0)-\mathbf{A}(t_i).\nonumber
\end{equation}
Here, $t_i=t_0+\tau$, and $\tau$ is the Coulomb-induced ionization time lag relative to the Coulomb-free ionization time $t_0$.
In \cite{Xie,Wang2020,Che2}, with the use of  a  Coulomb-modified SFA (MSFA) model
which is related to numerical solution of Coulomb-included Newton equation
for each SFA electron trajectory \cite{Lewenstein1995,Becker2002},
it has been shown that the introduction of this lag concept  into the SM mapping is able to qualitatively explain complex strong-field phenomena \cite{Xie,Wang2020,Che2}.
However, the strict definition and further quantitative description of this lag are far from being realized.
To verify the applicability of the above Coulomb-included mapping, two things need to be performed.
Firstly, one needs to give an analytical expression which can clearly define and exactly calculate the time lag $\tau$.
Secondly,  the observable deduced from the time lag $\tau$ with this mapping needs to be validated by experiments.
These are the main works in the paper.

We use the offset angle in PMD as the characteristic quantity to test the mapping $\mathbf{p}'=\textbf{v}(t_0)-\mathbf{A}(t_i)$.
The offset angle is related to the most probable route (MPR), for which the tunnel event occurs at the peak time $t_0$ of the laser field with $|E_x(t_0)|=E_0$ and $v_x(t_0)=0$.
The offset angle $\theta$ in our theory can be expressed as \cite{method}
\begin{equation}
\tan\theta=p'_x/p'_y= A_x(t_i)/(A_y(t_i)-v_y(t_0)).
\end{equation}
Equation (1) establishes the relation between the observable $\theta$ and the lag $\tau=t_i-t_0$.
When $\gamma\ll1$, with considering $|v_y(t_0)/A_y(t_0)|\ll1$, we also have $\mathbf{p}'\approx-\mathbf{A}(t_i)$. Then we have $\tan\theta\approx A_x(t_i)/A_y(t_i).$
Below, we will call the expression that neglects $v_y(t_0)$ `adiabatic Eq. (1)'.
Here, $\gamma=w\sqrt{2I_p}/E_0$ is the Keldysh parameter \cite{Keldysh}.

Figure 1 is plotted to give an intuitive picture for the lag and its relation with the offset angle. A sketch of the lag $\tau$ is presented in Fig. 1(a). The definition of the angle $\theta$ in PMD,
obtained through numerical solution \cite{Feit} of the time-dependent Schr\"{o}dinger equation (TDSE) for the He atom in two-dimensional (2D) cases  (see methods in \cite{method}), is indicated in Fig. 1(b).
This angle disappears in Fig. 1(c) of SFA simulations without $\tau$,
and is well reproduced in Fig. 1(d) with the proposed TRCM method which considers $\tau$ and will be introduced below.

\begin{figure}[t]
\begin{center}
\rotatebox{0}{\resizebox *{6.3cm}{6cm} {\includegraphics {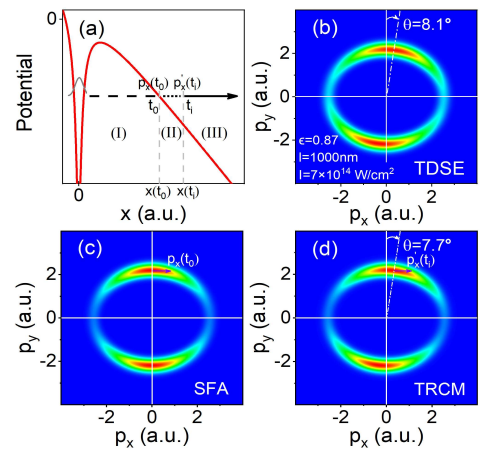}}}
\end{center}
\caption{
Sketch of the ionization time lag (i.e., the response time) for the MPR and its characterization in PMD.
When the electron exits the  barrier at
the peak time $t_0$ of the laser field (related to the drift momentum $p_x\equiv p_x(t_0)=0$),
it is not free immediately and a time lag $\tau=t_i-t_0$
between the exit time $t_0$ and the ionization time $t_i$ (related to $p'_x\equiv p'_x(t_i)\neq0$) emerges (a).
This lag reflects the response time of the electron to light in strong-laser-induced tunneling ionization.
It can be evaluated with Eq. (3) and can also be read from the offset angle $\theta$ in PMD (b)
with  Eq. (1). With the knowledge of response time,
the process of strong-field ionization can be divided into three steps of
Tunneling (I), Response (II) and Classic Motion (III).
These steps can be described with saddle-point, semiclassical  and SM theories, respectively,
raising a model termed as TRCM.
With TRCM, the Coulomb-free PMD of SFA (c) can be directly transited into the Coulomb-included one (d), in good agreement with TDSE (b)
and without the need of solving Newton equation including Coulomb force.
Laser parameters used are as shown.
}
\label{fig1}
\end{figure}

Next, we explore the analytical expression of $\tau$. This tunnel exit is generally not far away from the nucleus.
We assume that at the tunnel exit, the tunneling electron is still located at a quasi-bound state which approximately
agrees with the virial theorem. Semiclassical treatment of the quasi-bound state gives a velocity
$|v_{ix}|\approx\sqrt{|V(\mathbf{r}(t_0))|/n_f}$ which points to the nucleus and 
reflects the basic symmetry requirement of the Coulomb potential on the electric state.
Here, $n_f=2,3$ is the dimension of the single-electron system studied. 
A small period of time $\tau$ is then needed for the tunneling electron to obtain the opposite velocity  $-v_{ix}$ 
in order to break this basic symmetry and free itself. 
For MPR, this implies $E_0\tau\approx |v_{ix}|$. Then we can obtain the analytical expression of the lag (see methods in \cite{method})
\begin{equation}
\tau\approx\sqrt{|V(\mathbf{r}(t_0))|/n_f}/E_0.
\end{equation}
The above expression is one of the main results of this paper.
The derivation of this expression indeed reveals that the lag $\tau$
reflects the finite response time of the electronic wave function to a tunneling ionization event.
The corresponding response process occurs around the tunnel exit and arises from the strong interaction of the laser,
the electron and the atomic nucleus.
In particular,  in the mapping $\mathbf{p}'=\textbf{v}(t_0)-\mathbf{A}(t_i=t_0+\tau)$, the lag $\tau$
is the only time that describes the timescale of this three-body interaction.
Thus it quantifies the response time of the electron
to light in a tunneling ionization event characterized by this interaction.
It should also be stressed that the lag $\tau$ defined here encodes the significant effect of the
Coulomb potential during tunneling
and this near-nucleus Coulomb effect is described quantum mechanically with the virial theorem.
This is different from the MSFA where the Coulomb effect is considered classically after the tunneling electron exits the barrier.

\begin{figure}[t]
\begin{center}
\rotatebox{0}{\resizebox *{6.2cm}{5.5cm} {\includegraphics {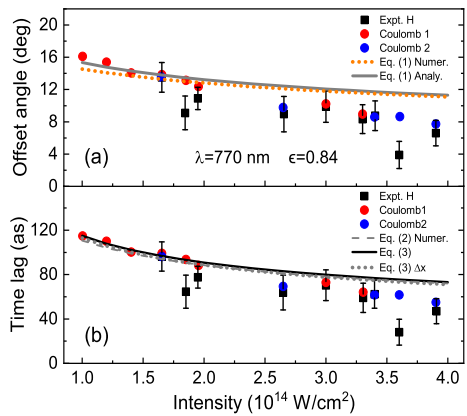}}}
\end{center}
\caption{Application to H for predicting the offset angle and the time lag.
Dots in (a) and (b): experimental  (black square) and 3D-TDSE  (red and blue circles) results  in \cite{Undurti}.
Lines in (a): predictions of Eq. (1) with $\tau$ and $v_y(t_0)$ evaluated using the numerical solution of SPE (orange dotted) or using the analytical expressions (gray solid).
Lines in (b): predictions of Eq. (2) with the exit position $x(t_0)$ evaluated using the numerical solution of SPE (gray dashed),
 predictions of Eq. (3) (black solid), and
predictions of Eq. (3) with the displacement correction $\triangle x\approx Z/(6I_p)$ to $x(t_0)$ (gray dotted).
Laser parameters used are as shown.}
\label{fig2}
\end{figure}

Let us further discuss the analytical treatment of the lag $\tau$.
In Eq. (2), the exit position $\mathbf{r}(t_0)$ can be evaluated with solving the saddle-point equation (SPE) ${[\mathbf{p}+\mathbf{A}(t_s)]^2}/{2}=-I_p$.
By neglecting the field $E_y(t)$ in solving SPE, we also have
$x(t_0)\approx (E_0/\omega^2)[\sqrt{\gamma^2+1}-1]$ and $y(t_0)\approx0$.
In the single-active electron approximation, the potential $V(\textbf{r})$ for a hydrogen-like atom has the form of $V(\textbf{r})=-Z/r$.
Then we have  \cite{method}
\begin{equation}
\tau\approx\sqrt{Z\omega^2/[n_fE_0^3(\sqrt{\gamma^2+1}-1)]}.
\end{equation}
Here, $Z$ is the effective charge. For real three-dimensional (3D) cases in experiments, the value of $Z$ can be evaluated with $Z=\sqrt{2I_p}$.
For TDSE, the value of $Z$ can be chosen as that used in simulations.
Equation (3) shows that the value of  $\tau$ is determined by the laser and atomic parameters of $E_0$, $\omega$ and $I_p$.
This value is about 100 attoseconds for general cases (see Fig. 2).

Once the lag $\tau$  is obtained with Eq. (2) or Eq. (3), we can evaluate the offset angle $\theta$ through Eq. (1)  at $t_i=t_0+\tau$.
We do so with two manners,  the exact one where we calculate the lag $\tau$ of Eq. (2) and the velocity $v_y(t_0)$
both with the numerical solution of  SPE (Eq. (1) Numer.),
and the approximate analytical one  where we calculate $\tau$ with Eq. (3) and $v_y(t_0)$ with  $v_y(t_0)=[\epsilon\sqrt{2I_p}/\text{arcsinh}(\gamma)-E_1/\omega]\sin\omega t_0$
obtained with neglecting the field $E_y(t)$ in solving SPE (Eq. (1) Analy.).
Note, the analytical one is applicable for MPR and for a small $\gamma$.
To validate our above discussions related to the response time $\tau$, next, we apply our theory to different targets and compare the observable $\theta$ deduced from $\tau$ with real and numerical experiments.

\begin{figure}[t]
\begin{center}
\rotatebox{0}{\resizebox *{7cm}{5cm} {\includegraphics {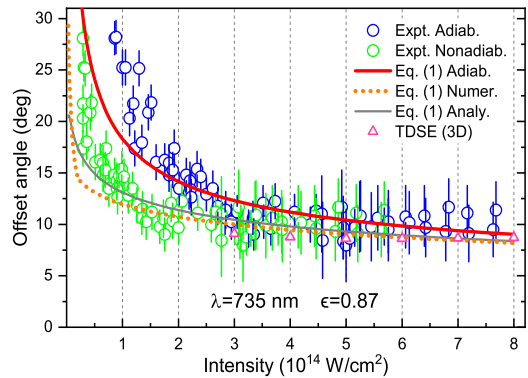}}}
\end{center}
\caption{Application to He for predicting the offset angle. Dots: experimental results with the adiabatic (blue circle) or
nonadiabatic (green circle) laser-intensity calibration in \cite{Boge}.
Red line: predictions of `adiabatic Eq. (1)' with $\tau$ calculated using Eq. (3) (red solid).
Orange and gray lines: same as in Fig. 2(a).
Magenta triangles:  3D-TDSE results.
Laser parameters used are as shown.}
\label{fig3}
\end{figure}

\emph{Application to H}.-We first apply our theory to the H atom with comparing to experimental and 3D-TDSE data in \cite{Undurti},
as shown in Fig. 2.
In our theory, the time lag $\tau$ is determined by laser and atomic parameters (Eq. (3)) and
the observable $\theta$ is deduced from the lag $\tau$ through the mapping Eq. (1). In \cite{Undurti},
the time delay $\tau$  is deduced from the measured angle $\theta$ with the relation $\theta\approx\omega\tau$.
For both cases of $\theta$ and $\tau$,  our theory predicts the decrease of the corresponding values  with the increase of laser intensity.
This decrease is  slower for high laser intensities than low ones.
These predicted phenomena are well verified by the experimental  data in \cite{Undurti}.
Quantitatively, our theory results agree with these data for $I\leq2\times10^{14}$ W/cm$^{2}$.
There is a difference of about two degrees or 10 attoseconds for higher intensities in Fig. 2(a) or Fig. 2(b).

With using the numerical solution of SPE to evaluate the exit position $\textbf{r}(t_0)$ in Eq. (2), or
using the expression of $x(t_i)=x(t_0)+\Delta x$ (also see Fig. 1(a) for the definition of $x(t_i)$),
which considers the displacement difference $\Delta x\approx v_{ix}^2/(2E_0)\approx Z/(6I_p)$,
to replace $x(t_0)$ in Eq. (3), the theory results in Fig. 2(b) become somewhat smaller and nearer to experimental results in \cite{Undurti}.
The difference between our results and experiments at high laser intensities may be due to the fact
that the ionization of H with $I_p=0.5$ a.u. is also strong for high intensities and this effect is not considered in our theory.

\emph{Application to He}.-Next, we apply our theory to the He atom with different treatments of the velocity $v_y(t_0)$ in Eq. (1),
and compare with related experimental data for the offset angle $\theta$ in \cite{Boge}, as shown in Fig. 3.
For $I\ge1.5\times10^{14}$ W/cm$^{2}$, the curve of `adiabatic Eq. (1)', which neglects  $v_y(t_0)$ and is related to the adiabatic intensity scaling in experiments, passes well through the corresponding experimental data.
For lower intensities, it deviates a bit from the  experimental result, but
can reproduce the trend of the experimental angle which increases rapidly for lower intensities.

The situation is similar for predictions of Eq. (1), which includes $v_y(t_0)$ and corresponds to the nonadiabatic intensity scaling in experiments.
Both the numerical and analytical curves of Eq. (1) agree well with  the nonadiabatic experimental data for $I\geq1\times10^{14}$ W/cm$^{2}$.
The remaining difference between theory and experiment for cases of low  intensities  results from the fact that the Coulomb effect plays a more important role in the momentum $\mathbf{p}'$ for these cases,
and this role is underestimated in our theory. Here, some 3D-TDSE results from our numerical experiments \cite{method} are also presented for predicting real experiments at higher laser intensities.

\begin{figure}[t]
\begin{center}
\rotatebox{0}{\resizebox *{8cm}{5.5cm} {\includegraphics {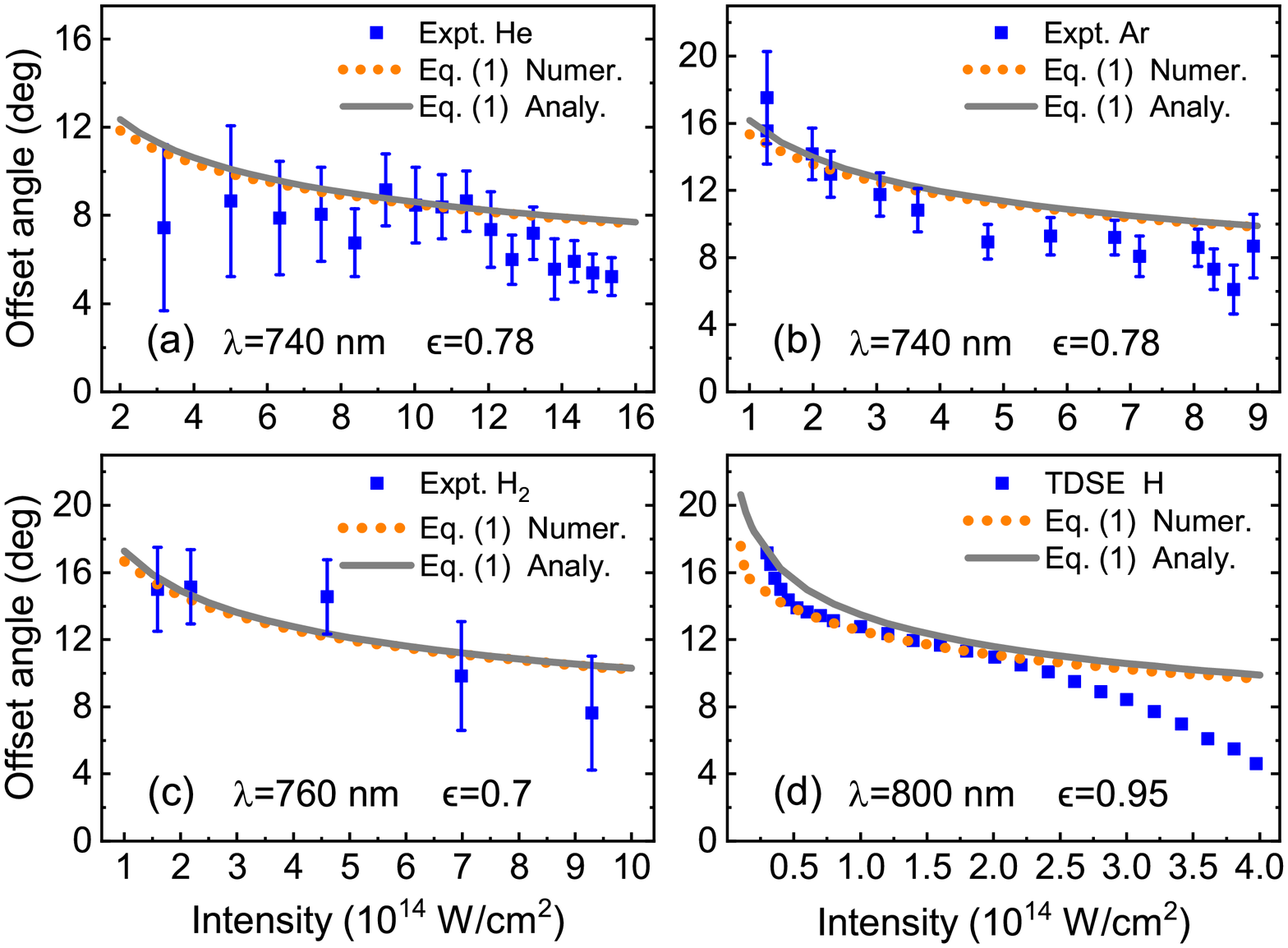}}}
\end{center}
\caption{Application to more cases for predicting the offset angle.
Blue square dots: experimental  results for He (a) and Ar (b) in \cite{Eckle3}
and for H$_2$ (c) in \cite{Quan}, and 3D-TDSE results (denoted with H2) for H (d) in \cite{Torlina}.
Lines in (a)-(d): same as in Fig. 2(a).
Laser parameters used are as shown.}
\label{fig4}
\end{figure}
\emph{Application to more cases}.-We have also applied our theory to more  targets and a wider range of laser parameters,
with comparing to experimental  and TDSE data for H \cite{Torlina}, Ar \cite{Eckle3} and H$_2$ \cite{Quan}, etc., as shown in Fig. 4.
In all cases,  our response-time theory manifested with Eqs. (1)-(3) well reproduces the main characteristics of relevant experimental or TDSE results. In particular, comparisons in Fig. 4(c) for H$_2$ with $R=1.4$ a.u. show that our theory is also applicable for molecules with a small internuclear distance $R$. Results in Fig. 4(d) for H are somewhat similar to those in Fig. 2(a) with theory curves agreeing with the TDSE one  from  $I=0.4\times10^{14}$ W/cm$^{2}$ to $I=2.5\times10^{14}$ W/cm$^{2}$. Extended comparisons between our theory predictions and TDSE simulations for
different laser wavelengthes and ionization potentials also support that our theory has the general applicability
and holds the essence of the physics behind the phenomena studied (see Figs. S1-S3 in \cite{method}).

\begin{figure}[t]
\begin{center}
\rotatebox{0}{\resizebox *{8cm}{5.5cm} {\includegraphics {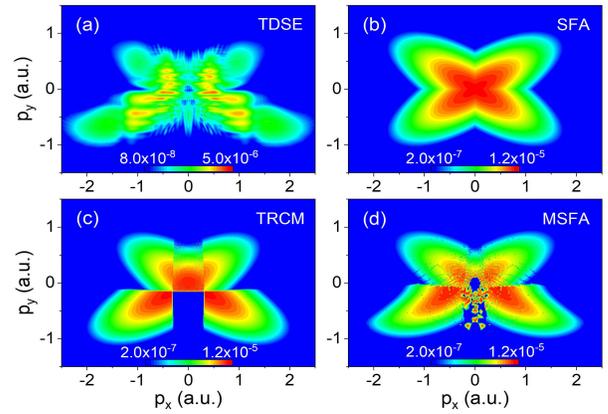}}}
\end{center}
\caption{Application to OTC laser fields.
Results here show PMDs of He obtained with 2D-TDSE (a), SFA (b), TRCM (c) and MSFA (d) in an OTC laser field of
$\mathbf{E}(t)=f(t)[\vec{\mathbf{e}}_{x}E_{x}(t)+\vec{\mathbf{e}}_{y}E_{y}(t)]$ with $E_{x}(t)=E_0\sin(\omega t)$ and $E_{y}(t)=\varsigma E_0\sin(2\omega t+\phi_0)$.
Laser parameters used are $I_x=5\times10^{14}$ W/cm$^2$ ($E_0\approx0.12$ a.u.), $\lambda_x=1000$ nm ($\omega\approx0.046$ a.u.),
$\varsigma=0.5$ and $\phi_0=\pi/2$.
The $\log_{10}$ scale is used here.}
\label{fig12}
\end{figure}
\emph{TRCM model}.-Methodologically, Eq. (2) is applicable for the MPR. For a general SFA electron trajectory ($\textbf{p},t_0$) with the amplitude $c(\textbf{p},t_0)$ \cite{Lewenstein1995},
we have $\tau\approx\sqrt{|V(\mathbf{r}(t_0))|/n_f}/E(t_0)$, where  $E(t_0)$ is the amplitude of the laser field. For the elliptical case,  $E(t_0)=\sqrt{(E_0\sin\omega t_0)^2+(E_1\cos\omega t_0)^2}$. Then using the mapping  $\mathbf{p}'=\textbf{v}(t_0)-\mathbf{A}(t_i=t_0+\tau)$, one can directly obtain the Coulomb-included PMD
with the drift momentum $\mathbf{p}'$ and the amplitude $c(\mathbf{p}',t_i)\equiv c(\textbf{p},t_0)$, without the need of solving Newton equation (see methods  in \cite{method}).
This theory, which can be called as TRCM (see Fig. 1(d)), provides a simple tool for
quantitative study of strong-field ionization dynamics  of atoms and small molecules with long-range Coulomb potential in diverse laser fields.

Figure 5 shows an application of TRCM to He in an orthogonal two-color (OTC) laser field.
The result of TDSE in Fig. 5(a) shows a butterfly-like structure with a remarkable up-down asymmetry.
This structure is absent in SFA results in Fig. 5(b),
where we solve  the SPE  to obtain the electron trajectory ($\textbf{p},t_0$) and the corresponding amplitude $c(\textbf{p},t_0)$ \cite{Lewenstein1995}.
This remarkable structure is  reproduced by the TRCM in Fig. 5(c).
Meanwhile, the MSFA results in Fig. 5(d), obtained with numerical solution of the Newton equation  for each SFA electron trajectory ($\textbf{p},t_0$) \cite{Xie}, also show a similar structure.

The remaining difference between results of TDSE and TRCM  is that the distribution around the origin in Fig. 5(a) shows small amplitudes, while that in Fig. 5(c) shows large amplitudes.
By comparison, the MSFA result in Fig. 5(d)  is similar to the TDSE one,
with showing small amplitudes around the origin. We therefore expect that the recapturing process,
which occurs when the rescattering electron approaches the nucleus, plays a dominating role in the
distribution of TDSE around the origin. This process is not considered in TRCM.

\emph{Discussions}.-The agreement between our theory and experiments supports the semiclassical response process around tunnel exit depicted in TRCM,
where quantum tunneling is time-free but a finite response time related to this semiclassical process is needed.
The response process
indicates a narrow time-space boundary between quantum and classic,
which is characterized by a time scale of $\tau\sim100$ attoseconds and a space scale of $\Delta x\sim0.3$ a.u..
It is interesting to further imagine the possible quantum-classic boundary in the case of photoemission induced by a weak laser field or related to single-photon transition.
Physically, the photoelectric effect is always related to the transition of the electron from a quantum-behavior-prevailing bound state  to a classic-behavior-dominating free state.
On the other hand, this agreement shows that it is possible to measure the electron-to-light response time in tunneling ionization unambiguously with the present ultrafast laser technology.
For example, for attoclock experiments, the peak time $t_0$ of the laser field timings the beginning of the response process and the ionization time $t_i$ of the electron timings the end.
These time information is encoded in the amplitude and momentum of the MPR related to the offset angle and can be retrieved with the mapping of Eq. (1).

It is worth noting that for the limit case of $\gamma^2\rightarrow0$, Eq. (3) can be approximated as $\tau\approx(\frac{3}{2}E_0\sqrt{2I_p})^{-\frac{1}{2}}$  for real 3D atoms.
Although the limit expression seems applicable only to tunneling ionization, it provides a simple tool for roughly evaluating the timescale of a photoemission process at present.
Further experiments for different targets and laser parameters are highly desired.

\emph{Conclusion}.-We have addressed the subtle issue of the response time of the electron inside an atom to light in strong-field tunneling ionization. 
A semiclassical theory has been developed to describe the response process and a simple mapping between the photoelectron momentum and the response time has been established.
We have shown that how the response time remarkably influences the observable and how it can be probed with the present ultrafast laser technology.
Our theory can be applied to different targets and to diverse forms of laser fields,
with providing a simple tool for
quantitatively explaining and predicting strong-field ultrafast phenomena.
In particular, our approach which treats the interaction time at the boundary between quantum and classic opens a perspective
for studying the response time of photoemission in other light-matter interactions.

We thank Y. F. He for discussions. This work was supported by the National Natural Science Foundation of China (Grant Nos. 12174239, 11904072),
and the Fundamental Research Funds for the Central Universities of China (Grant No. 2021TS089).

\appendix

\section{Methods}

\subsection{Numerical method}

The Hamiltonian of the He atom studied here has the  form of ${H}(t)=H_{0}+\mathbf{E}(t)\cdot \mathbf{r}$ (in atomic units of $\hbar=e=m_e=1$).
Here $H_{0}={\mathbf{{p}}^2}/{2}+V(\mathbf{r})$  is the field-free Hamiltonian and $V(\textbf{r})=-Z/\sqrt{r^2+\xi}$ is the Coulomb potential.
For 2D cases of $V(\textbf{r})=-Z/\sqrt{x^2+y^2+\xi}$, with the effective charge $Z=1.45$ and the soft-core parameter $\xi=0.5$, the ionization potential of the system reproduced here is $I_p=0.9$ a.u..
The term $\mathbf{E}(t)=f(t)[\vec{\mathbf{e}}_{x}E_{x}(t)+\vec{\mathbf{e}}_{y}E_{y}(t)]$ with $E_{x}(t)=E_0\sin(\omega t)$ and $E_{y}(t)=E_1\cos(\omega t)$  is the electric field of the elliptically-polarized laser field.
Here, the term $\vec{\mathbf{e}}_{x}$  ($\vec{\mathbf{e}}_{y}$) is the unit vector along the  $x$ ($y$) axis (i.e., the major (minor) axis of the polarization ellipse).
The term $\epsilon$ is the laser ellipticity, $\omega$ is the laser frequency, and $f(t)$ is the envelope function.
$E_0={E_L}/{\sqrt{1+\epsilon^2}}$, $E_1=\epsilon {E_L}/{\sqrt{1+\epsilon^2}}$,
and $E_L$ is the maximal laser amplitude related to the peak intensity $I$ of the laser pulse.
The value of $\epsilon$ used here is $\epsilon=0.87$.
We use  trapezoidally shaped laser pulses with a total duration of 15 optical cycles and linear ramps of three optical cycles.
The TDSE of $i\dot{\Psi}(t)=$H$(t)\Psi(t)$ is solved numerically using the spectral method \cite{Feit}.
We work with a grid size of $L_x\times L_y=409.6\times 409.6$ a.u..
The space steps used are $\triangle x=\triangle y=0.4$ a.u., and the time step is $\triangle t=0.05$ a.u..

To avoid the reflection of the electron wave packet from the boundary and obtain the momentum space wavefunction, the coordinate
space is split into the inner and the outer regions with ${\Psi}(t)={\Psi}_{in}(t)+{\Psi}_{out}(t)$, by multiplication using a mask function
$F(\mathbf{r})=F(x,y)=\cos^{1/2}[\pi(r_b-r_f)/(L_r-2r_f)]$ for $r_b\geq r_f$ and $F(x,y)=1$  for $r_b< r_f$.
Here, $r_b=\sqrt{x^2+y^2/\epsilon^2}$, $r_f=2.1x_q$ with $x_q=E_0/\omega^2$  and $L_r/2=r_f+50$ a.u. with $L_r\leq L_x$.
The above procedure considers the factors that the quiver amplitude of the ionized electron differs for different
laser parameters and for  $x$ and $y$ directions.
In the inner region, the wave function ${\Psi}_{in}(t)$ is propagated with the complete Hamiltonian $H(t)$. In the outer region, the time evolution of
the wave function ${\Psi}_{out}(t)$ is carried out in momentum space with the Hamiltonian of the free electron in the laser field.
The mask function is applied at each time  interval  of 0.5 a.u. and the obtained new fractions of the outer wave function are added
to the momentum-space wave function $\tilde{{\Psi}}_{out}(t)$ from which we obtain the PMD.
Then we find the local maxima of the PMD and the offset angle $\theta$ is obtained with a Gaussian fit of the angle distribution of local maxima.

For 3D cases of He, we have used the parameters of $Z=1.34$ and $\xi=0.071$ in the expression of $V(\textbf{r})=-Z/\sqrt{x^2+y^2+z^2+\xi}$.
The grid size used here is $L_x\times L_y\times L_z=358.4\times 358.4\times 51.2$ a.u. with $\triangle x=\triangle y=0.7$ a.u.
and $\triangle z=0.8$ a.u..
The value of $Z$ used here agrees with the relation of $Z=\sqrt{2I_p}$. The mask function used here is $F(\mathbf{r})=F_1(x,y)F_2(z)$.
The expression of $F_1(x,y)$ is similar to $F(x,y)$ used in 2D cases. The expression of $F_2(z)$ is $F_2(z)=\cos^{1/2}[\pi(|z|-r_z)/(L_z-2r_z)]$ for $|z|\geq r_z$ and $F_2(z)=1$  for $|z|< r_z$.
Here, $r_z=19.2$ a.u. is the absorbing boundary along the $z$ direction.
The numerical convergence is checked by using a finer grid.

\subsection{Analytical method}
In the part, we introduce the Coulomb-included strong-field model termed as TRCM, which arises from strong-field approximation (SFA)
\cite{Lewenstein1995} but considers the Coulomb effect \cite{MishaY,Goreslavski,yantm2010}.
This model puts an emphasis on the Coulomb potential near the nucleus where the quantum effect is strong
and assumes that the tunneling electron is located at a transition state possessing properties of both bound and continuum states.
Then it transfers the Coulomb effect into an ionization time lag, with establishing a definite
Coulomb-included mapping between photoelectron momentum and ionization time.
This mapping can be used to study the response time of the electron to light in strong-field ionization.

\subsubsection{Coulomb-free momentum-time mapping}
First, according to the SFA with the saddle-point method \cite{Lewenstein1995,Becker2002},
 the main contributions to a strong-field ionization event characterized by the photoelectron momentum $\textbf{p}$
 come from some specific electron trajectories, which agree with  the following  saddle-point equation
\begin{equation}
{[\mathbf{p}+\mathbf{A}(t_s)]^2}/{2}=-I_p. \tag{A.1}
\end{equation}
Here, $\textbf{A}(t)$ is the vector potential of the electric field  $\mathbf{E}(t)$.
The solution $t_s=t_{0}+it_{x}$ of the above equation is complex. The real part $t_0$ can be understood as the tunneling-out time.
Without considering the Coulomb potential, the tunneling-out time $t_0$ also amounts to the ionization time at which
the electron is free.
For $I_p=0$, one can return to the SM mapping relation between time and momentum. That is
\begin{equation}
\textbf{p}=-\textbf{A}(t_0). \tag{A.2}
\end{equation}

For a real atom with $I_p\neq0$, the SFA mapping relation between time and momentum can be written as
\begin{equation}
\textbf{p}=\textbf{v}(t_0)-\textbf{A}(t_0). \tag{A.3}
\end{equation}
Here, the term $\textbf{v}(t_{0})=\mathbf{p}+\textbf{A}(t_{0})$
denotes the exit velocity of the photoelectron at the exit position (i.e., the tunnel exit)
$\mathbf{r}_0\equiv\mathbf{r}(t_0)=Re(\int^{t_0}_{t_0+it_{x}}[\mathbf{p}+\mathbf{A}(t')]dt')$ \cite{yantm2010}.
This velocity reflects the basic quantum effect of tunneling.
The corresponding complex amplitude $c(\textbf{p},t_0)$
for the electron trajectory ($\textbf{p},t_0$)
can be written as $c(\textbf{p},t_0)\equiv c(\textbf{p},t_s)\sim e^{b}$.
Here, $b$ is the imaginary part of the quasiclassical action
$S(\textbf{p},t_s)=\int_{t_s}\{{[\textbf{p}+\textbf{A}(t'})]^2/2+I_p\}dt'$ with $t_s=t_0+it_x$ \cite{Lewenstein1995}.

\subsubsection{Coulomb-included momentum-time mapping}
The exit position can be roughly evaluated with $r_0\approx I_p/E_0$. For general laser and atomic parameters used in experiments,
such as the He atom exposed to a strong elliptical laser field with $I=5\times10^{14}$ W/cm$^{2}$ and $\varepsilon=0.87$,
the exit position is about $10$ a.u. away from the nucleus.
 Around this distance, the high-energy bound state of the field-free Hamiltonian H$_0$ has large probability amplitudes.
We therefore assume that for a real atom, around the tunnel exit $\mathbf{r}(t_0)$, the electron wave packet related to the tunneling electron
with the momentum $\textbf{p}$ is consisted of high-energy bound states. In other words, the tunneling electron is
still located at a quasi-bound state $\psi_b(\textbf{r})=\sum_n a_n|n\rangle$ at the time $t_0$.
Here, $|n\rangle$ is the bound eigenstate of H$_{0}$.
Such a state approximately agrees with  the virial theorem. That is $\langle\textbf{v}^2/2\rangle\approx-\langle V(\textbf{r})/2\rangle$.
In fact, $\langle\textbf{v}^2/2\rangle=-\langle V(\textbf{r})/2\rangle-\langle V(\textbf{r})/2\rangle_{n\neq m}$. Here,  $\langle V(\textbf{r})/2\rangle_{n\neq m}=\sum^{n\neq m}_{n,m}a_n^*a_m\langle n|V(\textbf{r})/2|m\rangle$
denotes the contributions of the off-diagonal terms
to the average potential energy. The contributions of different off-diagonal terms cancel each other.
In addition, the absolute amplitude of a certain off-diagonal term is generally  smaller than the corresponding diagonal one.
Therefore the amplitude of $\langle V(\textbf{r})/2\rangle_{n\neq m}$ is usually small in comparison with
$\langle V(\textbf{r})/2\rangle$ which includes both diagonal and off-diagonal contributions.

We continue with the idea of SFA. That is, after the tunneling electron exits the barrier at the time $t_0$, it can be treated as a free particle.
We assume that when the Coulomb potential is considered,
the bound wave packet $\psi_b(\textbf{r})$ related to the tunneling electron can also be treated
as a quasi-free particle with a velocity $\textbf{v}_i$ agreeing with $\textbf{v}_i^2/2=\langle\textbf{v}^2/2\rangle\approx-\langle V(\textbf{r})/2\rangle\approx-V(\textbf{r}_0)/2$ and $\textbf{v}_i=-|\textbf{v}_i|\textbf{r}_0/r_0$.
Namely, we assume that the kinetic energy $\textbf{v}_i^2/2$ of the quasi-free electron agrees with the virial theorem
but the  direction of the velocity $\textbf{v}_i$ induced by the Coulomb potential is contrary to the direction of the exit velocity $\textbf{v}(t_0)$ induced by the laser field.
To do so, we in fact introduce a quasi-free electron with a ``minus" kinetic energy.
Then according to the simple-man picture \cite{Corkum}, the quasi-free electron which exits the barrier at the time $t_0$
with a laser-induced exit velocity $\textbf{v}(t_0)$ and a Coulomb-induced one $\textbf{v}_i$ agrees with the following mapping relation
\begin{equation}
 \mathbf{p}'=\textbf{v}(t_{0})+\textbf{v}_i-\textbf{A}(t_0). \tag{A.4}
\end{equation}
Here, $\mathbf{p}'$ is the Coulomb-included drift momentum of the tunneling electron.
In the above expression, the velocity $\textbf{v}_i$ is introduced to describe
the effect of the Coulomb potential during the tunneling process when the electron is near the nucleus.
On the other hand, recent studies  showed that the Coulomb effect in strong-field ionization
also manifests itself as an ionization time lag in comparison with the SFA prediction \cite{Xie}. Considering this time lag effect,
we assume that at the time $t_i=t_0+\tau$ with a lag $\tau$ to $t_0$, the following relation holds.
That is $-\textbf{A}(t_i)\simeq\textbf{v}_i-\textbf{A}(t_0)$. Then we arrive at
\begin{equation}
 \mathbf{p}'=\textbf{v}(t_{0})-\textbf{A}(t_{i}).  \tag{A.5}
\end{equation}
The above expression shows that the Coulomb effect related to the presumed velocity $\textbf{v}_i$
induces a lag $\tau$ of the ionization time $t_i$ relative to the tunneling-out time $t_0$.
After the time $t_i$, the tunneling electron is free with 
the initial velocity $\textbf{v}(t_i)\equiv\textbf{v}(t_0)$. 
It is driven by only the laser field and the Coulomb potential is negligible. The latter assumption is also reasonable
since the Coulomb effect is more remarkable when the electron is near the nucleus than far away from the nucleus. 

With the above discussions, this lag $\tau$ can be further understood as the observable response time of the electron inside an atom to light 
in strong-laser-induced photoelectric effects.
Specifically, due to the existence of the Coulomb potential, the tunneling electron appearing at the tunnel exit
at the time $t_0$ is not free immediately. Instead,  under the action of laser field, 
a small period of time $\tau$ is needed for the tunneling-electron wave packet
to evolve from the transition state $\psi_b$ which contains both bound and continuum properties 
into an ionized state  which is Coulomb-free.  
Before the tunneling-out time $t_0$, the tunneling process described by SFA with saddle points is real-time free, 
and after the ionization time $t_i$, the Coulomb potential is also neglected, 
so the lag $\tau$ includes all the observable response time of the electron to light 
in strong three-body interaction between electron, nucleus and photon in our treatment.

It should also be stressed that equation (A.5) is applicable for cases where the rescattering effect plays a small role,
such as the case of the near-circular laser field.
We will return to this point later.

\subsubsection{Coulomb-included angle-time mapping}
In attoclock experiments, the offset angle is used as the characteristic quantity to deduce the time information
from PMD. This offset angle is defined by the part of PMD which is associated with the most probable route (MPR) and has the maximal amplitude.
This MPR is related to the electron trajectory ($\textbf{p},t_0$) with
the time $t_0$ corresponding to the peak time of the major-axis component $E_x(t)$ of the elliptical laser field. That is $|E_x(t_0)|=E_0$.
Some properties of MPR in the elliptical case are as follows.
The initial velocity $v_x(t_0)$ for MPR is zero, i.e., $v_x(t_0)=0$, and that of $v_y(t_0)$ has a nonzero value arising
from the nonadiabatic effect  \cite{Boge}.
When $\gamma\sim1$,  the  value of $|v_y(t_0)|$ is comparable to  $|A_y(t_0)|=E_1/\omega$, but for $\gamma\ll1$,  $|v_y(t_0)|/|A_y(t_0)|\ll1$. Here, $\gamma=w\sqrt{2I_p}/E_0$ is the Keldysh parameter \cite{Keldysh}.

Considering  Eq. (A.5) and the relation $v_x(t_0)=0$,
we can define the offset angle $\theta$  with  (Eq. (1) in the main text)
\begin{equation}
\tan\theta=p'_x/p'_y= A_x(t_i)/(A_y(t_i)-v_y(t_0))\nonumber.
\end{equation}
This expression also indicates the Coulomb-included mapping relation between the offset angle $\theta$ and the ionization time $t_i=t_0+\tau$.
Through this expression, one can deduce the lag $\tau$ with the offset angle obtained in experiments or TDSE simulations.
When $\gamma\ll1$,  the absolute value of $v_y(t_0)$ is also far smaller than that of $A_y(t_0)$,
the Coulomb-included mapping relation  of  Eq. (A.5) can be approximated as $\mathbf{p}'\approx-\mathbf{A}(t_i)$.
To do so, we in fact introduce the lag $\tau$ into the SM mapping relation $\mathbf{p}=-\mathbf{A}(t_0)$.
Then we have
\begin{equation}
\tan\theta\approx A_x(t_i)/A_y(t_i).  \tag{A.6}
\end{equation}
This expression can be understood as the adiabatic version of Eq. (1) as discussed in the main text.
It has been used in \cite{Che2} to deduce the lag $\tau$ of the asymmetric HeH$^+$ system and has been termed as Coulomb-calibrated attoclock (CCAC). Here, the theory description  is given.

\subsubsection{Response time and its expression}
Next, we explore the analytical expression of the lag $\tau$ for MPR.
By the relations of  $-\textbf{A}(t_i)=\textbf{v}_i-\textbf{A}(t_0)$ and
$\textbf{v}_i^2/2=n_f{v}_{ix}^2/2\approx-V(\textbf{r}_0)/2$, we have  $|{A}_x(t_i)-{A}_x(t_0)|\approx E_0\tau\approx|{v}_{ix}|\approx\sqrt{|V(\textbf{r}_0)|/n_f}\equiv\sqrt{|V(\textbf{r}(t_0))|/n_f}$. Here, $n_f=2,3$ is the dimension of
the single-electron system studied.
For actual cases as in experiments, $n_f=3$.
Then we obtain (Eq. (2) in the main text)
\begin{equation}
\tau\approx\sqrt{|V(\mathbf{r}(t_0))|/n_f}/E_0\nonumber.
\end{equation}
With neglecting the field $E_y(t)$ in solving Eq. (A.1), the exit position  $\textbf{r}(t_0)$
can be approximated as
$x(t_0)\approx (E_0/\omega^2)[\sqrt{\gamma^2+1}-1]$ and $y(t_0)\approx0$.
In the single-active electron approximation, the potential $V(\textbf{r})$
for a hydrogen-like atom has the form of $V(\textbf{r})=-Z/r$, where $Z$ is the effective charge. Then we obtain (Eq. (3) in the main text)
\begin{equation}
\tau\approx\sqrt{Z\omega^2/[n_fE_0^3(\sqrt{\gamma^2+1}-1)]}\nonumber.
\end{equation}
For real 3D cases such as in experiments, the value of $Z$ can be evaluated with $Z=\sqrt{2I_p}$.
For TDSE simulations, the value of $Z$ can be chosen as that used in calculations.
Equation (3) shows that the lag $\tau$ decreases with the increase of the laser amplitude $E_0$ and the laser wavelength $\lambda$ (the decrease of the laser frequency $\omega$) on the whole.
It also shows that in TDSE simulations, the value of $\tau$  is larger in 2D cases  than  3D ones (see Fig. S1).
For $\gamma\ll1$, we also have $x(t_0)\approx (I_p/E_0)[1-\gamma^2/4]$ and Eq. (3) can be further approximated as $\tau\approx\sqrt{Z/[n_fI_pE_0(1-\gamma^2/4)]}$.

It should be noted that Eq. (2) is applicable only for the long-range Coulomb potential. For a short-range potential,
$V(\textbf{r}(t_0))\rightarrow0$ and therefore $\tau\rightarrow0$.

\subsubsection{TRCM model}

Equation (2) is obtained for the specific electron trajectory of MPR in a strong elliptical laser field.
It can be extended to general SFA electron trajectories ($\textbf{p},t_0$) in different forms of laser fields. That is
\begin{equation}
\tau\approx\sqrt{|V(\mathbf{r}(t_0))|/n_f}/E(t_0).  \tag{A.7}
\end{equation}
Here, $E(t_0)=|\mathbf{E}(t_0)|$ is the amplitude of the laser electric field $\mathbf{E}(t)$ at the time $t_0$.
For the elliptically-polarized case, we have $E(t_0)=\sqrt{(E_0\sin\omega t_0)^2+(E_1\cos\omega t_0)^2}$.
Once the lag $\tau$  is obtained, using  Eq. (A.5), we can obtain the Coulomb-included
drift momentum $\mathbf{p}'=\textbf{v}(t_0)-\mathbf{A}(t_i)$ with $t_i=t_0+\tau$ and $\textbf{v}(t_0)=\textbf{p}+\mathbf{A}(t_0)$.
Assuming that the amplitude $c(\mathbf{p}',t_i)$ for the Coulomb-included electron trajectory ($\mathbf{p}',t_i$) is equivalent to
the corresponding amplitude $c(\textbf{p},t_0)\sim e^{b}$ for the SFA trajectory $(\textbf{p},t_0)$ \cite{Lewenstein1995}, we can obtain the Coulomb-included PMD directly from the SFA without the need of
solving Newton equation including both the electric force and the Coulomb force.
As the above theory naturally arises from the three-steps picture of Tunneling, Response and Classic Motion (TRCM) for strong-field ionization of  real atoms, depicted in Fig. 1 in the main text,
we would like to call it TRCM. This TRCM can be applied to various cases with different targets and diverse laser fields,
and provides a simple tool to explain and predict strong-field ionization phenomena.

It should be stressed that the TRCM assumes that after the ionization time $t_i$, the influence of the Coulomb potential
on the dynamics of the tunneling electron can be neglected. Therefore, this theory does not consider
the effects of rescattering and recapturing which are closely related to the Coulomb potential.
Generally, these effects play a small role for the trajectory away from the nucleus.
For the laser fields commonly used in attosecond experiments,
such as elliptical laser field and  orthogonal two-color laser field,
the TRCM works well except for some rescattering trajectories near to the nucleus. For these special cases, as shown in Fig. 5 in the main text,  the comparison between TRCM predictions and actual as well as numerical experiments also provides a method to identify these effects.
By further incorporating these effects into TRCM, we expect that the TRCM can also be used to explain related phenomena.

\setcounter{figure}{0}
\renewcommand{\thefigure}{S\arabic{figure}}
\captionsetup[figure]{name=Figure}
\begin{figure}[t]
\begin{center}
\rotatebox{0}{\resizebox *{8.5cm}{6.5cm} {\includegraphics {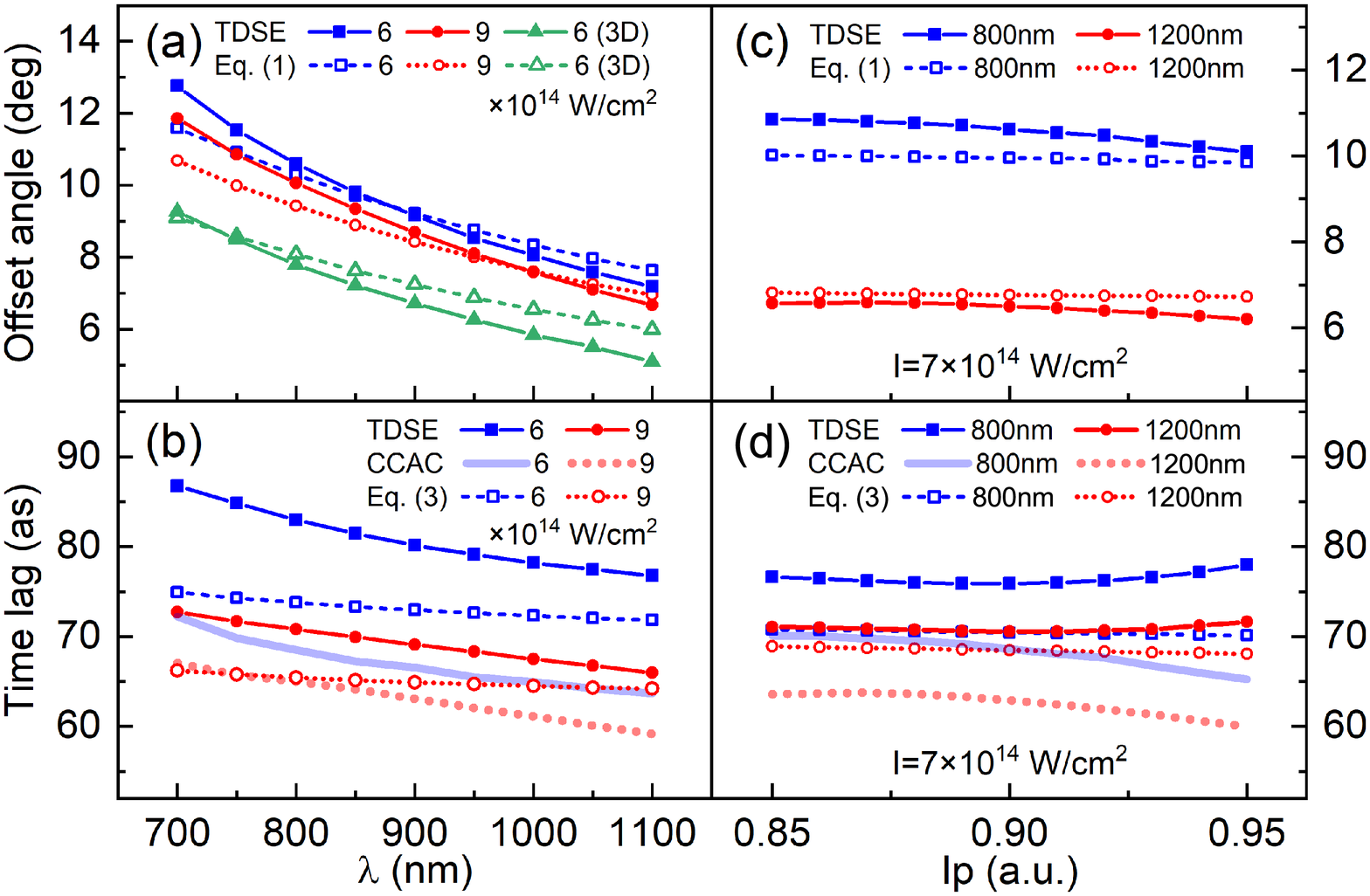}}}
\end{center}
\caption{Application to model He atom for predicting the offset angle and the time lag at different $\lambda$ and $I_p$.
The left (right) column shows results as a function of laser wavelength $\lambda$ (ionization potential $I_p$).
The first row:  predictions of  2D-TDSE and Eq. (1) for the offset angle, with $\tau$ and $v_y(t_0)$ in Eq. (1)
evaluated using the numerical solution of SPE.
The second row:  predictions of  2D-TDSE, CCAC and Eq. (3) for the lag.
For comparison, in (a), some 3D results (denoted with ``3D" to differentiate from 2D results) are also shown.
Laser parameters used are as shown and  $\epsilon=0.87$.}
\label{fig2}
\end{figure}

\begin{figure}[t]
\begin{center}
\rotatebox{0}{\resizebox *{8.5cm}{6.5cm} {\includegraphics {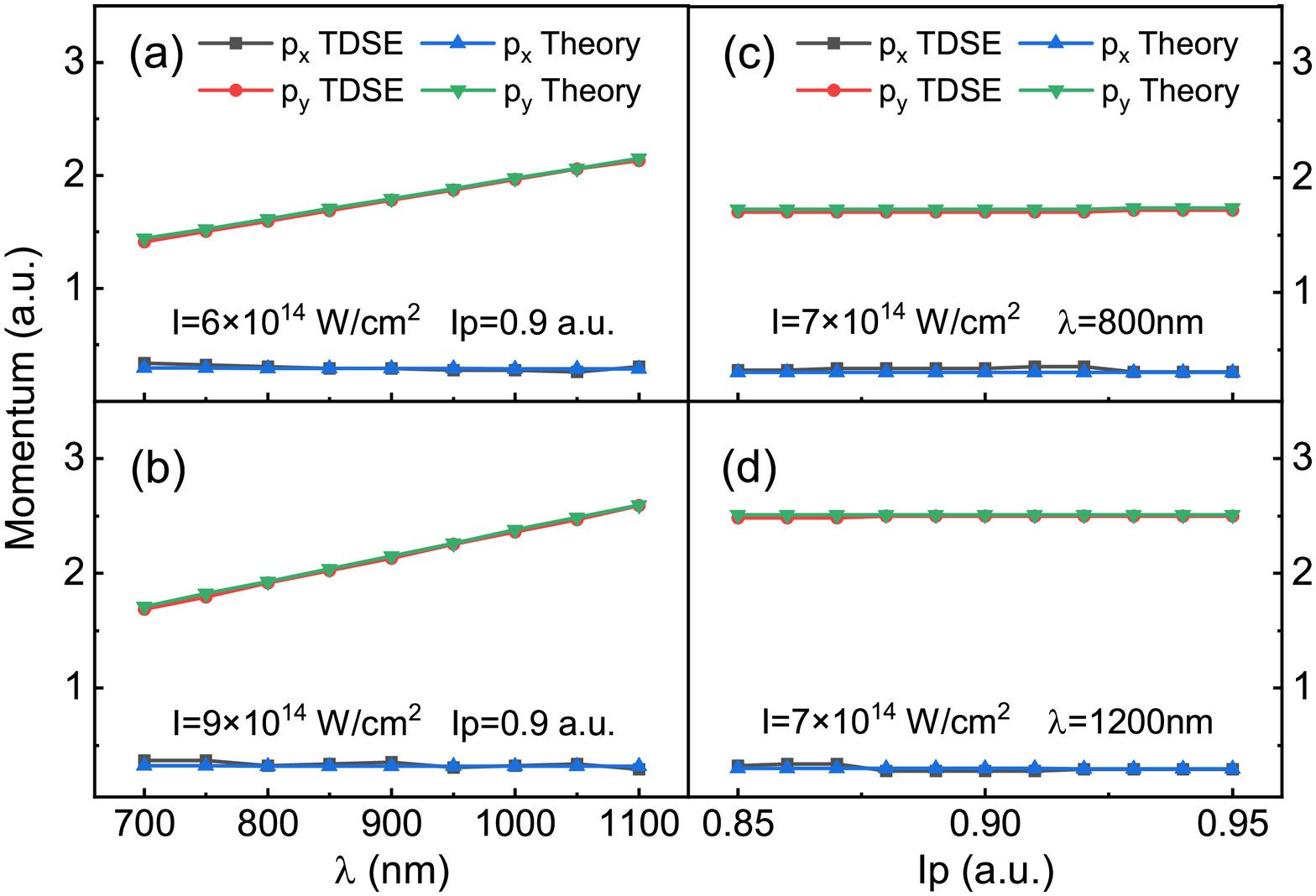}}}
\end{center}
\caption{Application to model He atom for predicting the drift momentum  associated with MPR at different $\lambda$ and $I_p$.
The results of 2D-TDSE are obtained with finding the drift momentum $(p_x,p_y)$ with the maximal amplitude in the PMD.
The theory results are obtained with the Coulomb-included mapping relation
$p_x=-{A}_x(t_i)$ and $p_y={v}_y(t_0)-{A}_y(t_i)$ at $|E_x(t_0)|=E_0$ and  $t_i=t_0+\tau$.
These values of $\tau$ and $v_y(t_0)$ are evaluated using the numerical solution of SPE. Laser parameters used are as shown and  $\epsilon=0.87$.
}
\label{fig11}
\end{figure}

\section{Extended comparisons for effects of $\lambda$ and $I_p$}

According to Eq. (3), the response time in tunneling ionization depends on the laser and atomic parameters of
$I$ ($E_0$), $\lambda$ ($\omega$) and $I_p$.
Present studies  mainly focus on the effect of laser intensity $I$ on the tunneling dynamics.
Here, we apply our theory to cases of different laser wavelengthes $\lambda$ and ionization potentials $I_p$, and compare
our theory predictions with extended TDSE simulations. Our TDSE calculations are first performed for 2D cases which allow us to explore a wide parameter region. Then we extend our considerations to 3D cases. Relevant results are first shown in Fig. S1.

In the first row of Fig. S1, we show the comparison for  the offset angle $\theta$.
One can observe that for a specific laser intensity, as increasing the laser wavelength, the offset angle of TDSE decreases.
In addition, at a certain wavelength, the TDSE offset angle is larger for the case of the lower laser intensity, as seen in Fig. S1(a).
The TDSE offset angle is not very  sensitive to the change of $I_p$ for the present parameter region, with a small decrease as increasing $I_p$,
as shown in Fig. S1(c). One can see that the theoretical predictions are very near to the TDSE ones and well  reproduce the remarkable parameter-dependent phenomena.
In Fig. S1(a), we also show some 3D TDSE results for predicting related experiments.

\begin{figure}[t]
\begin{center}
\rotatebox{0}{\resizebox *{8.5cm}{6.5cm} {\includegraphics {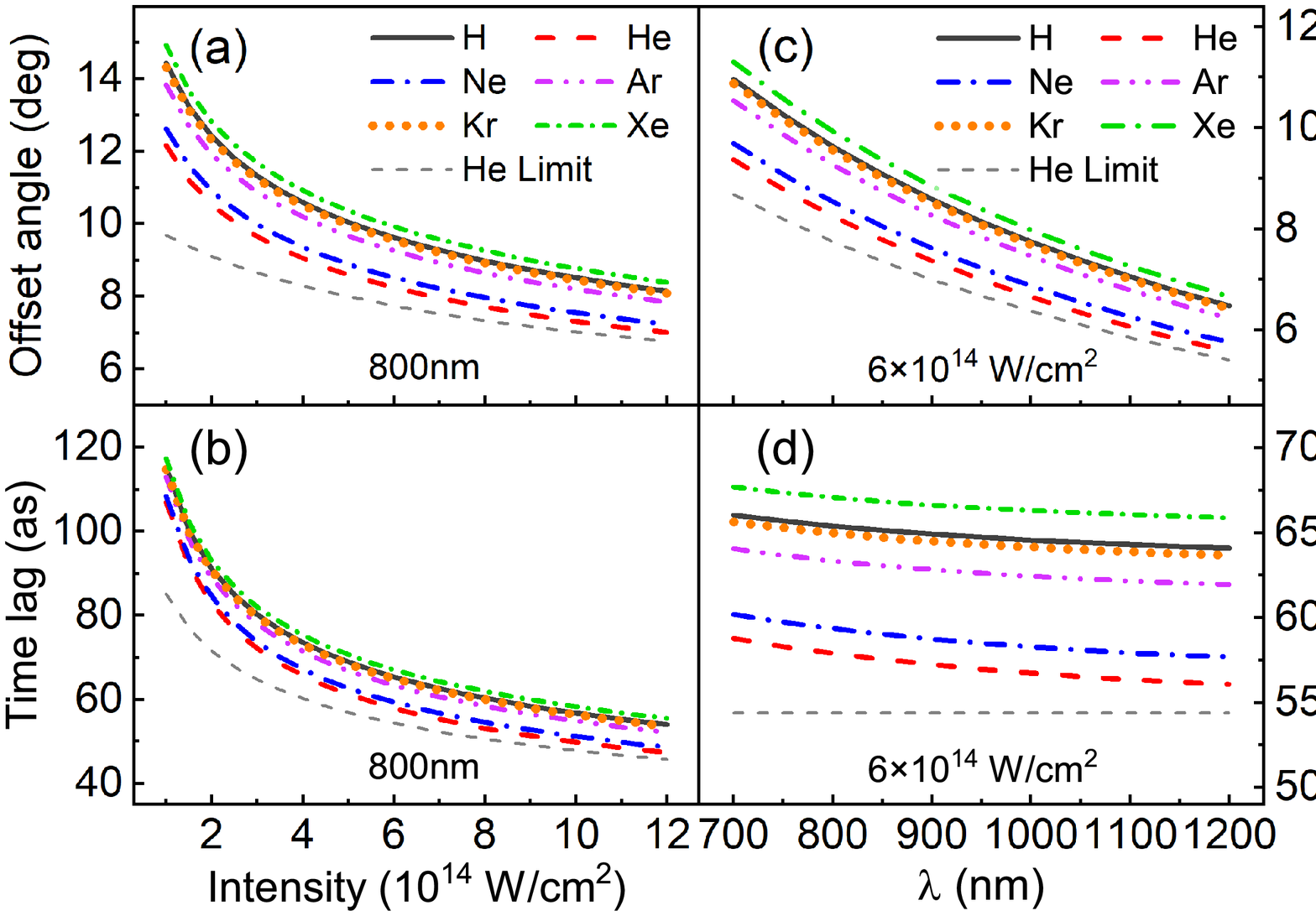}}}
\end{center}
\caption{Application to various real atoms for predicting the offset angle and the time lag at different $\lambda$ and $I$.
The time lags (the second row) of  atoms with different ionization potentials $I_p$ are obtained from Eq. (3) and the offset angles (the first row) are from  Eq. (1)
with $\tau$ and $v_y(t_0)$ evaluated using the analytical expressions.
The effective charges $Z$ used here agree with the relation of $Z=\sqrt{2I_p}$.
Results are presented as a function of the laser intensity $I$ (the first column) or the laser wavelength $\lambda$ (the second column).
In each panel,  the gray dashed line shows the limit result for the He atom, where the time lag $\tau$ is evaluated with the limit expression of $\tau\approx(\frac{3}{2}E_0\sqrt{2I_p})^{-\frac{1}{2}}$ at $\gamma^2\rightarrow0$.
Laser parameters used are as shown and  $\epsilon=0.87$.}
\label{fig10}
\end{figure}

Further comparisons for the lag $\tau$, obtained with Eq. (3), TDSE and CCAC, are presented in the second row of Fig. S1.
In CCAC, we first obtain the offset angle from the PMD of TDSE simulations. Then we obtain the time $t_i$ through adiabatic Eq. (1) (i.e., Eq. (A.6) in the method part) with $\theta\approx\arctan(A_x(t_i)/A_y(t_i))\sim\omega\tau$.
In TDSE, we first find the time $t_i$ which corresponds to the maximal value of the instantaneous ionization rate $P(t)=dI(t)/dt$.
Here, $I(t)=1-\sum_{m}|\langle m|\Psi(t)\rangle|^2$ is the instantaneous ionization yield,
 $|m\rangle$ is the bound eigenstate of H$_0=\textbf{p}^2/2+V(\textbf{r})$ and $|\Psi(t)\rangle$ is the TDSE wave function.
We only consider the first several bound eigenstates with $m=0,1,2...5$. The upper limit $m_u$ of $m$ is determined with the eigenenergy $E_{m_u+1}$ of the $(m_u+1)$th eigenstate agreeing with the semiclassical
analysis in Eq. (2). That is $E_{m_u+1}\approx V(\textbf{r}(t_0))+v_{ix}^2/2$.
Then the lag $\tau$ is obtained with $\tau=t_i-t_0$ at $|E_x(t_0)|=E_0$.
One can observe that for the broad parameter region, the difference for $\tau$ between results of Eq. (3) and TDSE or CCAC is near
to or smaller than 10 attoseconds.
This small difference between TDSE and Eq. (3) suggests the close correspondence for the definition of ionization between these two methods,
and that between CCAC  and Eq. (3) indicates that the lag $\tau$ (the response time) can be approximately evaluated
with the relation $\tau\sim\theta/\omega$, as discussed in Fig. 2 in the main text.

From Eq. (3), we also have $\tau\approx\sqrt{Z/(n_fI_pE_0)}$ when $\gamma^2\rightarrow0$. For real 3D cases, with $Z=\sqrt{2I_p}$,
we have $\tau\approx(\frac{3}{2}E_0\sqrt{2I_p})^{-\frac{1}{2}}$. This expression indicates that the value of $\tau$ is larger for smaller $I_p$ and smaller $E_0$ with the different scaling relations of
$\tau\sim I_p^{-\frac{1}{4}}$ and $\tau\sim E_0^{-\frac{1}{2}}$.
In practice, the ionization yield of the system depends strongly on $I_p$.  This limits the $I_p$-dependent comparisons in Fig. S1 to a small parameter region.
We will discuss the limit case of $\tau\approx(\frac{3}{2}E_0\sqrt{2I_p})^{-\frac{1}{2}}$ in details later.

More insights into roles of $\lambda$ and $I_p$ in the angle $\theta$ are obtained
when we compare the drift momentum $(p_x,p_y)$ of MPR between TDSE and theory predictions.
Relevant results are shown in Fig. S2 and the laser parameters used are as in Fig. S1.
Firstly, the  predictions of our theory  agree well with the TDSE ones both for $p_x$ and $p_y$.
Secondly, when fixing the laser intensity and increasing the laser wavelength (the left column), the value of $p_x$ almost does not change and the value of $p_y$ increases.
Therefore, it is the wavelength dependence of $p_y$ that mainly contributes to the wavelength dependence of the offset angle here.
Thirdly, for the present parameter region, both the values of $p_x$ and $p_y$ are insensitive to the small change of the ionization potential.
As a result, the offset angle is also insensitive to $I_p$. The results shed light on $\lambda$ and $I_p$-dependent phenomena in Fig. S1.
Note, the TDSE results presented here are obtained with simply finding the  peak of PMD and therefore show somewhat small fluctuations.

In Fig. S3, we show the predictions of $\theta$ and  $\tau$ by our theory for real atoms with diverse $I_p$,
calculated with the approximate analytical expressions of Eq. (3) and
$v_y(t_0)=[\epsilon\sqrt{2I_p}/\text{arcsinh}(\gamma)-E_1/\omega]\sin\omega t_0$ for $\tau$ and $v_y(t_0)$, respectively.
Firstly, when we fix the laser wavelength and increase the laser intensity (the first column), or with the contrary manipulation (the second column),
the calculated offset angles and the time lags of the targets both decrease, but the lag decreases slowly with the increase of the wavelength.
In particular, in all of cases, for fixed laser parameters, the calculated angles and lags are larger for atoms with smaller $I_p$.
This phenomenon can be understood with considering the limit case of  $\tau\approx(\frac{3}{2}E_0\sqrt{2I_p})^{-\frac{1}{2}}$ at $\gamma^2\rightarrow0$. This expression  shows that the lag $\tau$ decreases with increasing $I_p$. For the elliptical laser field  with high ellipticity, the approximation  $\theta\sim\omega\tau$ also holds.
Therefore, as changing $I_p$, the offset angle $\theta$ behaves similarly to the lag $\tau$.
As a case, in each panel of Fig. S3, we show the corresponding limit result for He. In comparison with general analytical results, the limit results are remarkably lower for cases of lower laser intensities
and shorter laser wavelengths corresponding to larger values of the parameter $\gamma=\omega\sqrt{2I_p}/E_0$, but approach the analytical ones for cases of small  $\gamma$.
As in the limit case, the value of  $\tau\approx(\frac{3}{2}E_0\sqrt{2I_p})^{-\frac{1}{2}}$ does not depend on $\omega$, the lag $\tau$ generally has a weak dependence on laser wavelength for smaller values of the parameter $\gamma$.
Because of $\theta\sim\omega\tau$, the offset angle $\theta$ decreases as the laser wavelength increases.

It should be stressed that in experiments, the $I_p$-dependent phenomena discussed above can change, since the ionization probability of an atom  depends strongly on its ionization potential.

\end{document}